\documentclass[pre,showpacs,twocolumn,superscriptaddress,preprintnumbers]{revtex4}
\usepackage{epsfig} 
\usepackage{color} 
\usepackage[T1]{fontenc}
\usepackage[latin1]{inputenc} 
\usepackage[english]{babel}
\usepackage{graphicx}

\newcommand {\ti} \textit

\newcommand {\avem} [2] {\langle \overline{#1}_{#2} \rangle}
\newcommand {\avemq} [2] {\langle \overline{#1}{#2} \rangle}
\newcommand {\ave} [1] {\langle #1 \rangle}
\newcommand {\comas} [1] {\textquoteleft #1 \textquoteright}

\begin{document}

\title{Exploring complex networks by means of adaptive walkers}
\bigskip
\author{Luce~Prignano}
\affiliation{Departament de F\'{\i}sica Fonamental, Universitat de Barcelona
08028 Barcelona, Spain}
\author{Yamir Moreno}
\affiliation{Institute for Biocomputation and Physics of Complex Systems (BIFI), University of Zaragoza, Zaragoza 50009, Spain}
\affiliation{Department of Theoretical Physics, University of Zaragoza, Zaragoza 50009, Spain}
\affiliation{Complex Networks and Systems Lagrange Lab, Institute for
   Scientific Interchange, Torino, Italy}
\author{Albert D\'{\i}az-Guilera}
\affiliation{Departament de F\'{\i}sica Fonamental, Universitat de Barcelona 08028 Barcelona, Spain}
\affiliation{Institute for Biocomputation and Physics of Complex Systems (BIFI), University of Zaragoza, Zaragoza 50009, Spain}

\date{\today}
\begin{abstract}
Finding efficient algorithms to explore large networks with the aim of recovering information about their structure
is an open problem. Here, we investigate this challenge by proposing a model in which random walkers with previously assigned home nodes navigate through the network during a fixed amount of time. We consider that the exploration is successful if the walker gets the information gathered back home, otherwise, no data is retrieved. Consequently, at each time step, the walkers, with some probability, have the choice to either go backward approaching their home or go farther away. We show that there is an optimal solution to this problem in terms of the average information retrieved and the degree of the home nodes and design an adaptive strategy based on the behavior of the random walker. Finally, we compare different strategies that emerge from the model in the context of network reconstruction. Our results could be useful for the discovery of unknown connections in large scale networks.
\end{abstract} 


\maketitle

\section{Introduction}
\label{section1}

During the last decades, much scientific interest has been devoted to 
the characterization and modeling of many natural and artificial systems that exhibit so-called emergent behavior. These systems, referred to as complex systems, are suitably described through their networks of contacts, that is, in terms of nodes (representing the system's components) and edges (standing for their interactions), which allows to catch their essential features in a simple and general representation. Complex networks \cite{newmanrev,physrep,caldarelli,dgm08} have therefore become an important, largely used, framework for the understanding of both dynamical and topological aspects of systems such as the brain \cite{booksporns}, protein-protein interaction networks \cite{bocahandbook}, Internet and the WWW \cite{book1}.  

In the meanwhile, it has also become clear that many of the mentioned networks, particularly those which are described by a power law degree distribution $P(k)\sim k^{-\gamma}$ (scale-free networks \cite{newmanrev,physrep,caldarelli,dgm08}),
are only partially known. Think, for instance, in online social networks like Facebook or Twitter, which are made up of millions of heterogeneous and non-identical nodes. In such large networks,  a complete map is hardly available and difficult to get \cite{facebook}. Thereby, providing efficient tools for their exploration has become a crucial challenge. In general, network features are discovered by means of algorithms based on search and traffic routing \cite{gdvca02,egm05,sclts07}. In many cases, the latter can be performed by means of moving "agents", which explore the topological space and recover information. Nonetheless, it is still a key issue the investigation and characterization of the efficiency of different strategies \cite{s-j05,fct07,pb12} as far as the quality and quantity of information gathered are concerned. 

On the other hand, it has also been shown that local topological metrics, like the degree of a node, greatly affect dynamical properties of complex networks. This is the case of immunization algorithms, which are more effective the larger the degree of the vaccinated node is \cite{cha04}. As a matter of fact, one of the best strategies is to immunize a neighbor of a randomly chosen node instead of the node itself. This is because a randomly chosen node has degree $k$, while a neighbor would have degree $k$ with probability $kP(k)$. Another striking example closely related to the problem here addressed in which the degree of the nodes determines dynamical properties is the scaling law characterizing flow fluctuations in complex networks \cite{flowba,flowba2,flowalex,mglm08}. Admittedly, the mean traffic $\langle f \rangle$ and its standard deviation $\sigma$ can be related through the simple scaling form $\sigma \sim \langle f \rangle^{\alpha}$ \cite{flowba,flowba2,flowalex}. However, the latter relation, which was previously thought to be universal with $\alpha$ being between $1/2$ and $1$, is not satisfied for all values of $k$, i.e., the exponent is not universal and depends, among other factors, on the degree of the nodes \cite{mglm08}.

In this paper, we address the problem of network exploration from the point of view of a single node from which an agent is sent through the network in order to collect \ti{information}, henceforth understood as the fraction of nodes visited when the walker gets back home. Our aim is to find out an optimal strategy to maximize both the number of visited nodes and the chance to meet again the starting point, independently of which is the choice for the latter. To this end, we consider an arbitrary (heterogeneous) network of $N$ nodes and a single agent (explorer or walker) initially located on a given node (home-node), and let it move during a time frame $T$, the walker's lifetime. Every time the agent comes back to the starting point, all the nodes it has visited until that moment are marked as \ti{visited} and the total information gathered is updated. Obviously, it could also be possible to send several agents at once, but it has been demonstrated for several similar situations \cite{INRIA} that increasing the number of walkers (and reducing their lifetime proportionally) does not produce better results. Consequently, we focus on the performance of single agents.

The most important novelty of our proposal is that the agents are not markovian random walkers, nor a modified version of random walks' dynamics in which additional rules (for instance, preferential or self-avoiding random walks \cite{fct07,vbf11}) are introduced. Indeed, we introduce a parameter $q$ which governs how likely it is for a walker, at each time step, to go forward or backward (with respect to the walker's home). Thus, by changing the value of this parameter, the two probabilities can be tuned and hence different strategies are defined. In one limiting case, the walkers will tend to move back home, whereas in the other limiting setting, they will tend to move away from home. In between these two asymptotic behaviors, we recover a classical random walk, for which all directions are equally probable. We explore different strategies and their dependencies with both the degree of the home nodes and the walkers' lifetimes. Moreover, we show that it is possible to built up an adaptive algorithm whose efficiency in terms of the information gathered and the quality of the reconstructed network is, in general, the best. 

The rest of the paper is organized as follows. Section\ \ref{section2} introduces the model which is characterized in Sections\ \ref{section3}-\ref{section4}. Our proposal for an adaptive strategy is presented in Section\ \ref{section5}. In Section\ \ref{section6} we present the application of the algorithms previously discussed to the reconstruction of the degree distribution. Finally, the last section (Sec.\ \ref{section7}) is devoted to round off the paper. 

\section{Baseline model of walkers}
\label{section2}

Let us first discuss a baseline model in which a given set of walkers explore the network starting from a home node. As previously discussed, in order to collect the results of walkers' exploration, they should go back home. Therefore, we introduce two probabilities when the walker is at a given node, provided it has tracked the information about the
path followed from the home-node to the current position. These two probabilities correspond to the forward (F) and backwards (B) motion along the already tracked path and read, respectively, as:
\begin{eqnarray}
P_F(k_i) &=& q^2(k_i - 1) / [1 + q^2(k_i -1 )] \label{pf},\\
\nonumber\\
P_B(k_i) &=& 1 / [1 + q^2(k_i -1)] \label{pb},
\end{eqnarray}
where the label $i$ indicates the node that the explorer is going to leave 
and $k_i$ is its degree. These equations stand for every step whenever the agent is not in the starting node $-$ the home, $h$ $-$. In the latter case, i.e., while at home, it can only go forward, thus at that position we have $P_F^h=1$ and $P_B^h=0$.

\begin{figure}[t]
\begin{center}
\includegraphics[width=.95\columnwidth]{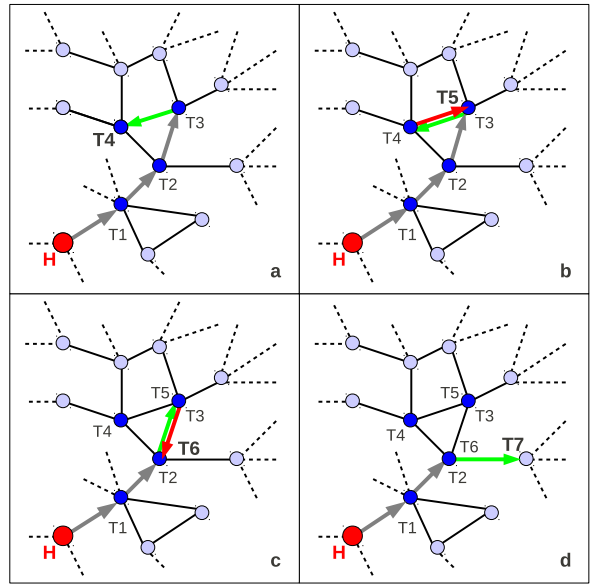}
\caption{(Color online). Example of the motion of an agent: 4 snapshots taken at 4 sequential time steps. The red node labeled with "H" is the home-node and the nodes that have been visited are colored in blue. Panel (a): T4. Grey arrows stand for previous steps forward, while the green arrow stands for the last one. Panel (b): T5. The agent takes a step backward (red arrow) reverting and removing from its "memory" the last step forward. Panel (c): T6. At this point it is the step taken at time instant T3 that has to be regarded as the last step forward (green arrow); the walker takes another step backward (red arrow) reverting it. Panel (d): T7. The walker takes a step forward toward a new node.}
\label{fig0}
\end{center}
\end{figure}

From Eqs.\ (\ref{pf})-(\ref{pb}), we recover the pure random walk (without any bias, i.e., all possible directions are equally probable) for $q=1$. For very large values 
of the parameter $q$, no backward step is allowed. Consequently the explorers can get back to their starting node only by chance, through a different path, not being \ti{aware} that they are coming back, but being able to recognize where they are (at home). Conversely, when $q$  goes to $0$, after the first move, no more steps forward are allowed. Therefore, only the first neighbors of the starting node can be explored. We also consider that the walker's lifetime is $T$ steps, which represents the time allowed for the network exploration before the dynamics stops. We define the information gathered as the fraction of nodes marked as \ti{visited} after $T$ time steps: $I=V/N$, where $V$ is the number of visited nodes and $N$ is the size of the network. Moreover, if the agent is not at home at time $T$, the new nodes visited after its last return to the home node are not computed in $V$ (i.e., we consider that only the information brought counts).

We first discuss the expected behavior of $I$ at the two limiting values of $q$ (very high or very small). On one hand, for very low $q$ values only the nearest neighbors are visited and hence $I$ will be small independently of $T$. On the other hand, for very large values of $q$ the walkers only return to home by chance, being the search also inefficient provided the exploration time is not very large (see next section). Then, if we fix the total number of steps we can expect that the information collected will have a maximum as a function of $q$. Therefore, there should exist, for any given network, a precise value $q^*(T)$ such that, if we average over all the possible choices of the home-node and over many realizations of the dynamical exploration, the mean information $\langle \overline{I}(q^*)\rangle$ is maximal. In other words, there is no other value $q'$ for which $\langle \overline{I}(q')\rangle > \langle \overline{I}(q^*)\rangle$, where \comas{$\,\langle\, \cdot\, \rangle$} stands for the mean performed over all the nodes in the network and \comas{$\,\overline{\,\cdot\,}$} for the average over many realizations.

The previous analysis indicates that the best efficiency in terms of maximal recovery of information can only be obtained for two values of $q^*$. In the next section, we explore the dependency of $I$ on the network properties (as given by the degree of the home node) and walkers' lifetimes. Admittedly, when this time is very long ($T \gg N$) we should expect to recover most information by setting $q^*\to \infty$. However, even if this is the best choice on average, it might not be the case when the home of the walker is at a low degree node. On the other hand, for shorter searching times, a value of $q=q^*<1$ gives almost the same performance for $I$, but this time the results are independent of the degree of the home node and $\langle \overline{I}(q^*)\rangle$ is a global maximum $-$ the caveat is that $q^*$ cannot be known \ti{a priori}.

\section{Characterizing the performance of the walkers}
\label{section3}

In this section we study the dependency between the information gathered by an agent and $q$, for different choices of the home-node and for different values of the walkers' lifetimes $T$. 
Hereafter we will use as a benchmark a scale free network of $N=10^4$ nodes and mean degree $\langle k\rangle=10$ generated by the uncorrelated configuration model \cite{cbps05}. We however note that all results reported are valid for any network with a power-law degree distribution provided that it does not have a tree-like topology. Actually, the only relevant difference in the case of a tree-like network is that we will observe a different behavior for large values of $q$. This is because leaves would make very difficult for a walker to come back through a different path making their performance very poor, even for very large values of $T$ and for very large degrees of the home nodes.

\begin{figure}[t]
\begin{center}
\includegraphics[width=\columnwidth]{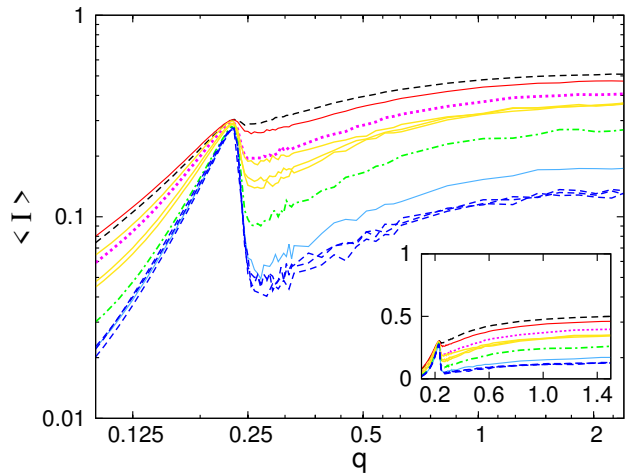}
\caption{(Color online) Information $\overline{I}$ (averaged over 3000 realizations) gathered by a walker during a searching time $T=10000$ as a function of the $q$ parameter. Each curve refers to a different home-node and different colors and line styles refer to different degrees of the starting node. From the top to bottom: $k=100$ (dashed black line), $k=54$ (solid red line), $k=30$ (dotted purple line), $k=22$ (solid yellow line), $k=13$ (dot-dashed green line), $k=7$ (solid light blue), $k=5$ (dashed blue line). In the inset: the same quantity in a linear scale.}
\label{fig1a}
\end{center}
\end{figure}

In Fig.\,\ref{fig1a} the information $\langle I\rangle$ is plotted as a function of $q$ for several home-nodes and a searching duration of $T=N=10000$ steps. 
As it is clearly shown, starting from small values of the parameter $q$, $\langle I\rangle$ initially increases but soon afterwards there is an abrupt decay to give way to a new increase as $q$ grows further. For very large values of $q$, the information gathered saturates to an asymptotic value. Interestingly enough, as seen in the figure, the amount of information gathered for both very small values of $q$ and when $q\gg1$, as well as the size of the abrupt decay, depend on the degree of the node from which the walker started the exploration. However, there exists a universal value of $q=q_p$ at which almost all curves corresponding to different degrees of the home node collapse, i.e., there is a local maximum which is roughly independent of the connectivity of the home node. Nevertheless,  whether this point is also a global maximum for $I(q)$ or just a local one depends on the degree of the initial node. Indeed, when the home-node is highly connected, for this searching duration, an agent performs better for $q\to \infty$, but if this is not the case, $q_p$ gives the optimum efficiency.

\begin{figure}[t]
\begin{center}
\includegraphics[width=\columnwidth]{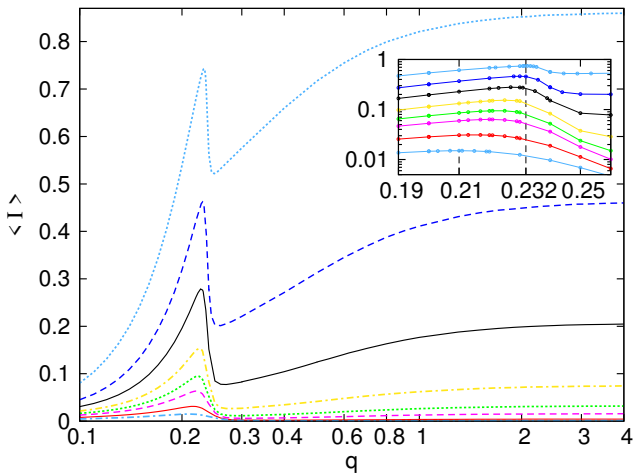}
\caption{(Color online) Mean Information $\avemq{I}{}$ gathered by a walker performing its search starting from any home-node during a time lag of $T$ steps, as a function of the $q$ parameter. The mean is performed over all the nodes in the network and averaging over 100 realizations for each. Different colors and line styles refer to different durations of the searching. From the bottom to the top: $T=500$ (light blue dot-dashed line), $1000$ (red solid line), $2000$ (dashed purple line), $3000$ (dotted green line), $5000$ (dot-dashed yellow line), $10000$ (solid black line), $20000$ (dashed blue line) and $50000$ (dotted light blue line). In the inset: zoom of the peak. Notice that $q_p$ displays a small shift increasing $T$, up to $T=10000$ when it reaches an asymptotic value.}
\label{fig2a}
\end{center}
\end{figure}

In Fig.\,\ref{fig2a} we plot the same quantity as in Fig.\,\ref{fig1a} but averaged over all the possible home-nodes (then the dependency with the degree washes out) and considering different lifetimes $T$. The figure makes it more clear that at $q=q_p$ the value of $\avem{I}{}$ is a global maximum unless $T$ is many times larger than the network size $N$. This definitively means that if we are interested in the information an agent may gather for a very long searching time, what we have to do is to set $q\gg1$. Otherwise, if we are interested in more realistic situations where there can be limitations on the duration of the exploration (for instance, due to energy constraints), the best choice would be to set $q=q_p$. The latter option has a caveat, however:  the precise value of $q_p$ depends in an unknown way on the topological features of the underlying network. Nevertheless, one can obtain useful insights into the problem by inspecting how the behavior of a walker changes when $q$ varies.

Looking more carefully to the results plotted in Fig.\,\ref{fig2a}, one can distinguish three regions that qualitatively correspond to the three distinct behaviors of the walker. In the first one, for $q<q_p$, $\avem{I}{}$ monotonously increases as a function of $q$; in the second one, $\avem{I}{}$ experiences an abrupt decay; whereas the third region shows that $\avem{I}{}$ starts to increase again, until it saturates to a value that depends on $T$. It is easy to realize that the first increase corresponds to small enough values of $q$. In this region, the walker moves just a few hops away home and consequently it takes only a few steps to get back home. The larger the value of $q$ is, the longer the mean path covered by the walker will be. Since for very small values of $q$ the exploration is local, the relevance of the home-node degree is very high (see Fig.\,\ref{fig1a}). Then, increasing $q$, we are allowing the walker to explore farther nodes, that is to collect new information, and the initial differences due to the degree of the home-node become progressively smaller. At $q=q_p$ they have almost vanished.

In the second region, for $q$ slightly larger than $q_p$, the walker often gets lost and its performance is, on average, less efficient. In other words, the explorer wastes an important fraction of the lifetime $T$ gathering information that it will not be able to bring back home before the time is over. The precise value at which this start to occurs is slightly affected by the duration of the exploration, as shown in the inset of Fig.\,\ref{fig2a}. This can be explained as a combination of two factors. On the one hand, to increase $q$ means to increase the number of nodes visited, but also the risk to get \comas{lost}. Indeed, if an agent is performing a long trip and it is going to bring a lot of information back home, when the searching time is suddenly over, the loss is big. On the other hand, the very first trips are those that provides the largest fraction of new information since the majority of nodes are being visited for the first time. Thus, getting lost after a couples of returns causes a much worse loss than if the same happens after a few round trips. Again it is a matter of balance and the optimum value $q_p$ is smaller when the lifetime is shorter. The second region ends at a value of $q$ for which the previous balance is the worst possible one, thus giving raise to another increase, which marks the start of the third region. Here, for even larger values of $q$, it begins to be quite frequent that, wandering across the network almost randomly, the explorer returns to its home-node through a different path just by chance. This new behavior entails a new increasing of $\avem{I}{}$ due to the fact that this kind of random returns start to balance the inefficiency of the walkers that get lost. The likelihood of these events increases with $q$ and it is maximum when $q\to \infty$, that is, when $P_B=0$ at each time step.

The previous dependency of $\avem{I}{}$ on the walker's lifetime $T$ defines two optimal values for $q$, either $\avemq{I}{(q)}$ takes its maximum value at $q*=q_p$ or at $q*=\infty$. However, we stress again that for $q\gg 1$, the walker gets back home by chance (recall that for these values of $q$ the backward probability $P_B=0$). Consequently the asymptotic values of $\avem{I}{}$ in the $q=\infty$ limit strongly depends on the degree of the home nodes (see Fig.\,\ref{fig1a}). Therefore, setting $q*=q_p$ could be a better choice even when $T$ is large enough. In order to be able to take advantage of the agents' behavior at $q_p$, we need to characterize deeper the transition that occurs for that value of the parameter. To this end, in the next section we focus on the behavior of some dynamical quantities which display a relevant change around $q_p$. 

\section{Exploration mechanisms and estimation of $q_p$}
\label{section4}

\begin{figure}[t]
\begin{center}
\includegraphics[width=.95\columnwidth]{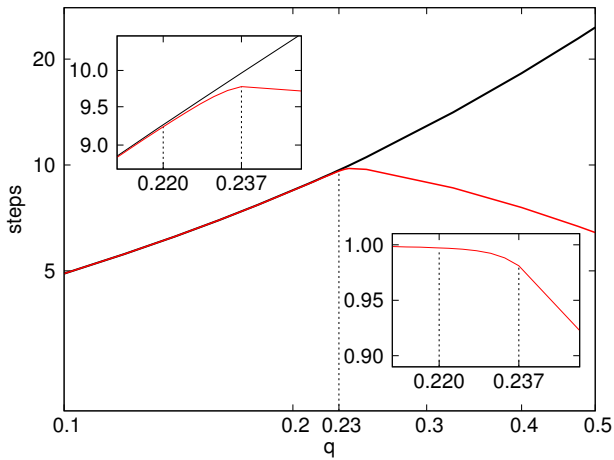}
\caption{(Color online) Mean maximum number of consecutive steps forward $\avem{S}{F}$ (black line) and backward $\avem{S}{B}$ (red line) taken by a walker during a time lag of $T=10000$ steps. The mean is performed over all the nodes and averaging over $2000$ realizations for each of them.
In the first inset (above): zoom around the value of $q$ at which the two curves get apart. In the second one (below): the ratio $\avem{S}{B}/\avem{S}{F}$ in the same range. According to the arguments discussed in this section, the peak should lie between the to values indicated with the dashed lines ($q_p \in [0.220\,;0.237]$) and this is in a good agreement with what we can observe in Fig.\,\ref{fig2a}.}
\label{fig3a}
\end{center}
\end{figure}

In Fig.\,\ref{fig3a} we plot the average maximum number of sequential steps backward (\,$\avem{S}{B}$\,) and forward (\,$\avem{S}{F}$\,) that a walker takes in a time lag $T=10000$ as a function of $q$. These two quantities, estimated by averaging over many realizations and over all the possible home-nodes, give a useful picture of the transition between the first and the second regimes previously described. They initially increase together, then $\avem{S}{B}$ start increasing slower than $\avem{S}{F}$, it reaches a maximum and start decreasing, asymptotically going to zero. 
Notice that for small $q$ the value of $\avem{S}{F}$ is small. Consequently, $\avem{S}{B}$ is bounded (even if $P_B(k)\sim 1\ \forall k$) since, when an agent is back to its home, no more steps backward can be taken. The value of $q$ for which $\avem{S}{B}$ and $\avem{S}{F}$ take the maximum value before getting apart roughly corresponds to $q_p$. It is when the walker goes as far as possible from its starting point, being still able to come back on its own steps. Increasing $q$ a little bit further provokes that the number of steps forward exceeds that of steps backward and the home-node is not recovered any more, so that the searching efficiency rapidly decreases. This phenomenology helps us to find out an heuristic definition for the peak. It is indeed possible to state that $q_p$ is the precise value of $q$ for which a walker is allowed to take enough steps forward to be able to visit a large region of the network, but at the same time it is also allowed to take enough steps backward so as to return to its home not by chance. 

\begin{figure}[tb]
\begin{center}
\includegraphics[width=.9\columnwidth]{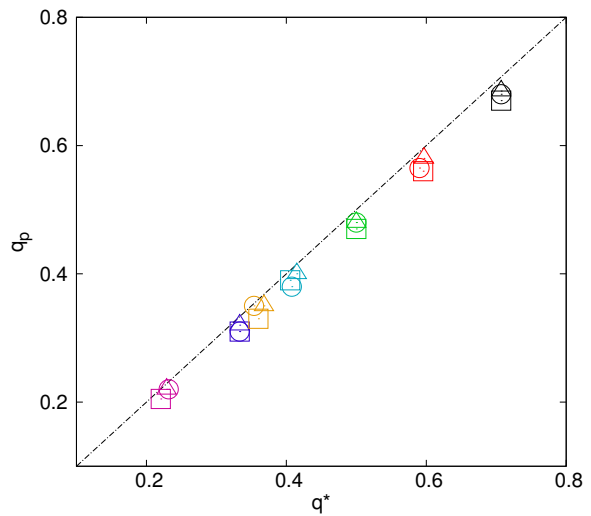}
\caption{(Color online). Estimator (\ref{qstar}) against the measured value of $q_p$ for several networks. All the networks have been generated by the uncorrelated configuration model changing the exponent $\gamma$ of the degree distribution together with the maximum degree $k_{max}$ and the minimum degree $k_{min}$, varying from quite heterogeneous topologies to random regular graphs. Different symbols stand for different network sizes: squares for $N=1000$, circles for $N=2000$, and triangles for $N=5000$. Each group of symbols corresponds to a given set of parameters. From the smallest value of $q_p$ we have: \{$\gamma$=1.5, $k_{max}$=50, $k_{min}$=2\}  (purple); random regular network with $k=10$ (blue); \{$\gamma$=2.5, $k_{max}$=50, $k_{min}$=2\}  (yellow); \{$\gamma$=2.5, $k_{max}$=30, $k_{min}$=2\}  (light blue); random regular with $k$=5 (green); \{$\gamma$=0.1, $k_{max}$=5, $k_{min}$=2\}  (red); random regular with $k$=3 (black). The values of $q_p$ have been measured for an exploring time $T$=$N$.}
\label{qstar-qp}
\end{center}
\end{figure}

Admittedly, it is possible to translate this heuristic statement into a quantitative condition starting from one simple observation. There exists, for any $k$, a value of $q$ such that $P_F(k)=P_B(k)$ and from Eqs.\ (\ref{pf})-(\ref{pb}) we know that this value is $q(k)=1/\sqrt{k-1}$. If $q=q(k_{max})$ it is guaranteed that $P_F\leq P_B$ $\forall k$, so the mean path is short and the explorer will come back to home very often. If $q=q(k_{min})$, the situation is the opposite, $P_F\geq P_B$ $\forall k$, so for the agent it is very difficult to recover its home. Therefore, the conclusion is that the peak lies between these two extremal values. A reasonable estimation could be obtained by imposing that $P_F/P_B=1$ on average while an explorer walks around. At each time step, the probability that a walker is on a node of degree $k$ is $P_w=kp(k)/\ave{k}$, where $p(k)$ is the degree distribution of the considered network. Hence, this condition can be rewritten as
\begin{equation}
\int_k \frac{P_F(k)}{P_B(k)}\frac{kp(k)}{\ave{k}}dk=q^2\int_k \frac{(k-1)kp(k)}{\ave{k}}dk=1,
\end{equation}
thus we obtain the estimator
\begin{equation}
q^*=\frac{1}{\sqrt{\ave{k^2}/\ave{k}-1}}.
\label{qstar}
\end{equation}
In Fig.\,\ref{qstar-qp} we have plotted $q^*$ against $q_p$ for several networks of different sizes and different topologies (degree distributions) finding a very good agreement in all of the considered cases. It can be confirmed that the precise value of $q_p$ only depends on the first and second moment of the degree distribution, while no explicit dependence on the network size can be observed, at least for finite $N$.

\begin{figure}[th]
\begin{center}
\includegraphics[width=\columnwidth]{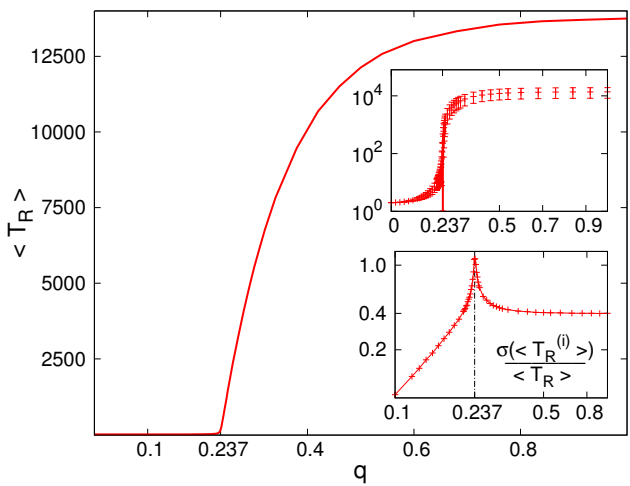}
\caption{Mean time (\,$\avem{T}{R}$\,) that a walker starting from any node in the network needs to come back to its home-node as a function of $q$. The average is performed over all the nodes, and for each of them over $5000$ realizations. In the first inset (above): on the left, the same quantity represented in a log scale with error bars (standard deviation among home-nodes); in the second one (below): the relative standard deviation of the values $\overline{T}_R^{(i)}$.}
\label{fig4a}
\end{center}
\end{figure}

In order to complete this phenomenological picture it can be useful to look at another quantity strictly related with what we have said in the previous paragraphs. 
In Fig.\,\ref{fig4a} we plot the mean time that a walker needs to come back to its home-node (\,$\avem{T}{R}$\,) as a function of $q$. In this case we did not set a lifetime. We then let $N_w$ walkers wander through the network starting from a given node. Each time an agent recovers its home it is not allowed to leave it anymore and the duration of the trip is recorded. We wait until every walker has come back and calculate the average return time $\overline{T}_R^{(i)}$, where $i \in [1,N]$ stands for the considered home-node. Finally we average over all the possible starting nodes. What we obtain is a curve that closely resembles that of the order parameter in a second order phase transition , with the critical point located slightly above $q_p$. Furthermore, if we look at the dispersion of the values of $T_R^{(i)}$ we recover a behavior quite similar to that of the susceptibility, i.e., a divergence at the critical point \cite{bfdn92}. Actually, the divergence takes place very close to $q_p$, but even closer to the value of $q$ for which the average number of consecutive backward steps is maximum (see the inset in Fig.\,\ref{fig3a}). With a slightly larger value of $q$, the return time starts to rapidly increase, with a corresponding abrupt increment in the dispersion. This indicates that, even if the trip duration uses to be small, sometimes - with a probability that increases by increasing $q$ - the agent needs a very long time in order to reach his/her home-nose. Thus, a lot of information is lost if the process is stopped before the explorer is able to complete its last journey. Again, this scenario corroborates the intuition that $q_p$  is the maximum value of $q$ able to guarantee that the walker will not get lost. 

\section{The adaptive strategy}
\label{section5}
For any given network we are now able to predict where the peak is located, given the first and the second moments of the degree distribution. However, we are interested in developing a searching strategy that can be useful when we have no information at all about the underlying topology.
In this section, we are going to set up an adaptive algorithm aimed at optimizing the performance of an agent exploring a heterogeneous network in a number of steps $T$ that is equal to (or less than) the number of nodes $N$. The basic idea is simple. We have a walker and a value of $q$ associated to it. We let it wander and when it is at home again we evaluate the contribution of this last round trip to the information gathered until that moment and, if necessary, the value of $q$ is modified. In order to build up such an algorithm, three main elements are needed.
The first one is an appropriate quantitative way to evaluate the performance of the agents. The second one is a criterion to decide whether or not $q$ would be modified. Finally, the adaptive rule applies whenever the choice is to change the value of $q$. This third element is an algorithm able to connect what the agent has learned about the network until its last return, the efficiency of its performance and the current value of $q$ in order to provide a new, more suitable, value for the parameter. 

Let us start with the first element. Since the aim of the exploration is to collect the maximum amount of information in a fixed time frame, to be \ti{efficient} means to visit as many new nodes as possible per unit of time (step). The final efficiency of a searching process can thus be defined as $E=I/T$. This definition can be expressed as a function of the number of round trips. If we indicate with $t_r$ the time of the $r$-$th$ return of the explorer ($0<t_1<t_2<\dots<T$), we have $E(t_r)=V(t_r)/t_r$, where $V(t_r)$ stands for the number of visited nodes after $t_r$ steps ($V(t_r)N=I(t_r)$). It is also possible to measure the efficiency of a single trip as $e_r=[V(t_r)-V(t_{r-1})]/(t_r-t_{r-1})]$, but this is not a very useful procedure as $e_r$ is very noisy. Therefore, in order to compare the performance at time $t_r$ with that at time $t_{r-1}$ it is better to consider the efficiency variation $\Delta E(t_r)=E(t_r)-E(t_{r-1})$. Hence, a good criterion to decide whether a change of $q$ is needed is $\Delta E(t_r)<0$. 

Notice that if we start with a small value of $q$, the number of steps forward and backward will be the same (see Fig.\,\ref{fig3a}) and the explorer will pass on each visited node at least two times. 
Therefore the first return time $t_1$ will be twice the number of steps ahead that the walker was allowed to take. The maximum number of different nodes that the agent may have visited during its trip is therefore equal to the number of steps it took forward. This happens whenever the walker does not cross each link more than twice (forward and backward). Thus, the efficiency has an upper bound, $E\leq 1/2$, that can be easily reached for any small value of $q$ when the explorer performs its first trip. In particular, for $q=0$ we surely have $E(t_1)=1/2$ since only one step forward is allowed and the agent will visit one node in two time steps. Consequently, we expect $E(t_r)$ to start from a value very close to $1/2$ and then necessarily decreases. Hence, changing $q$ has the effect of decelerating the decay of $E(t)$, or at most, to make $E(t)$ reach a stationary value. 

\begin{figure}[ht]
\begin{center}
\includegraphics[width=\columnwidth]{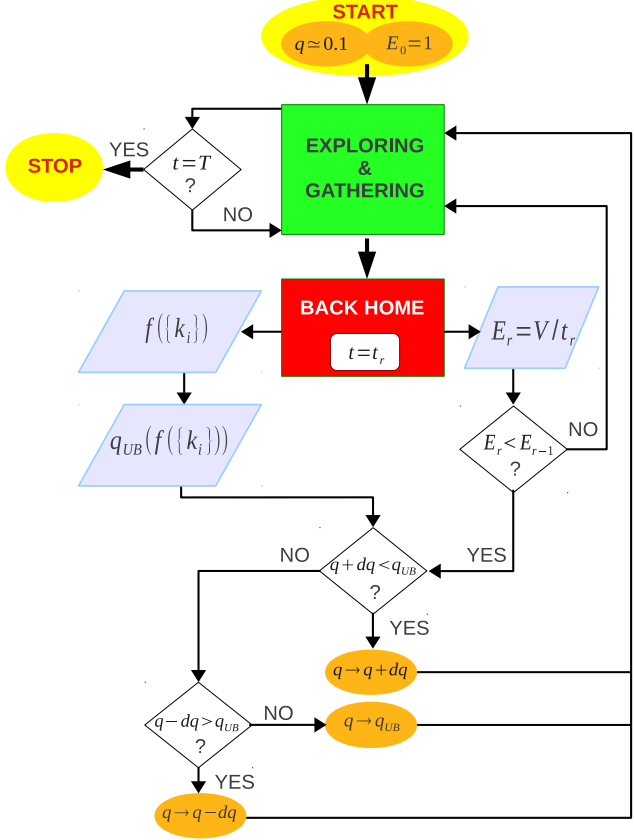}
\caption{(Color online). The figure represents, in a \ti{flowchart} a possible implementation of the algorithm described in the main text for the adaptive strategy. Here the efficiency $E$ takes the initial value $E_0=1$. The notation is simplified in order to make the diagram easier to read: $E(t_r)$ is indicated as $E_r$ and $V(t_r)$ just as $V$. }
\label{fig5a}
\end{center}
\end{figure}

When $q$ is varied, we should also take special care in not letting the agent to get lost.
For this reason, since the real value of $q_p$ is unknown, we need to start from a very small value of $q$ in order to ensure that $q\leq q_p$. Then, we want to let $q$ increase in a controlled way. Hence, we need to fix an upper bound for $q$ based on the information the agent is recovering about the degree of the visited nodes. A first, very simple election could be to use the estimator of $q_p$ provided by Eq.\,(\ref{qstar}). One can replace the probabilities $P_w(i)=k_ip(k_i)/\ave{k}$ with the visit frequencies $F_w(i)=n_i/t$, where $i$ is the node index, $n_i$ the number of times the walker has visited that node, and $t$ the elapsed time (number of steps). Thus we obtain the empirical estimator
\begin{equation}
q^*_e=\frac{1}{\sqrt{\sum _i^{V_t} [k_iF_i(k_i)] -1}},
\end{equation}
where $V_t$ is the number of visited nodes at time $t$.
Obviously, when $t\to \infty$ this empirical estimator is equal to the estimator (\ref{qstar}). The problem, however, is that this is not an upper bound. Actually, the degree distribution the walker recovers during the first trips is very noisy and $q^*_e$ fluctuates a lot. In some cases, it takes values quite smaller than the real $q_p$. This would prevent the explorer from increasing $q$ trapping him/her in the neighborhood of the starting node. Therefore, we need a quantity that satisfies the following requirements:
\begin{enumerate}
\item[a.] it has to be less noisy than $q^*_e$;
\item[b.] it has to take values smaller than $q_p$ very unlikely;
\item[c.] when evaluated over the whole network, its value has to be close to that of $q_p$;
\item[d.] it has to be the same as $q_p$ and $q^*_e$ when we consider a homogeneous network.
\end{enumerate} 
To satisfy the first requirement, we need to avoid to use the frequencies $F_i$ taking into account all the visited nodes with the same weight, regardless of how many times they have been visited. So we are looking for an appropriate function $f(\{k_i\})$ of the degrees of the visited nodes, such that $q_{UB}=1/\sqrt{f(\{k_i\})-1}$.
We propose the following expression that satisfies all the requirements:
\begin{eqnarray*}
f(\{k_i\})=\sqrt{\ave{k^2}(t)}=\sqrt{\sum_{i\in V_t}k_i^2/V_t}.
\end{eqnarray*}
Notice that in general $\sqrt{\ave{k^2}}\geq \ave{k^2}/\ave{k}$, where the equality holds in the case homogeneous networks.
Therefore, we have
\begin{equation}
q_{UP}=\left[\sqrt{\ave{k^2}(t)}-1\right]^{-1/2}.
\end{equation}
With all the previous remarks, the adaptive algorithm can be formulated as follows (see Fig.\ \ref{fig5a}):
\begin{itemize}
\item[1)] Set $q\sim 0.1$ and let the agent perform its first round trip.
\item[2)] Calculate $E(t_1)=V(t_1)/t_1$ and let the agent perform another trip.
\item[3)] Calculate the new value of the efficiency and check if $\Delta E(t_2)=E(t_1)-E(t_2)<0$. If it is not the case, let the agent explore again, until the condition $\Delta E(t_r)<0$ is satisfied.
\item[4)] Calculate $f(\{k_i\})$=$\sqrt{\ave{k^2}(t_r)}$ and then $q_{UB}(f(\{k_i\}))$.
\item[5)] Check if $q+dq<q_{UB}$, where $dq$ is a small positive quantity (in general $dq=0.01$ is a good choice).
\item[6)] If the condition (5) is satisfied, update the value of $q$ adding $dq$: $q \to q+dq$.
\item[7)] If the condition (5) is not satisfied, but $q<q_{UB}<q+dq$, update the value of $q$ so that $q\to q_{UB}$.
\item[8)] If $q>q_{UB}$ then:
\begin{itemize} 
\item[8a)] if $q-dq < q_{UB}$ then $q \to q_{UB}$, 
\item[8b)] if $q-dq > q_{UB}$ then $q \to q-dq$.
\end{itemize}
\end{itemize}

\begin{figure}[tbh]
\begin{center}
\includegraphics[width=\columnwidth]{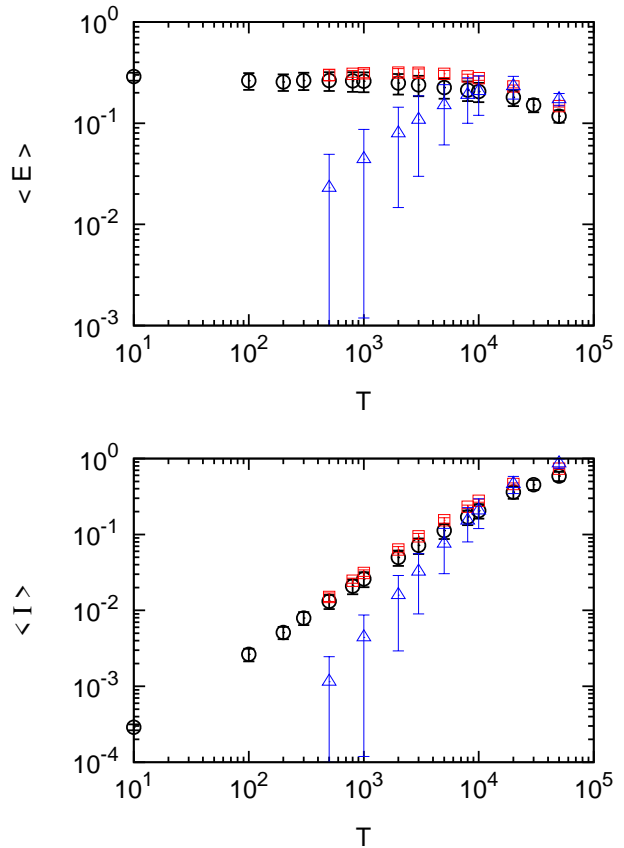}
\caption{(Color online) Top panel: the total efficiency $\avem{E}{}$=$\avem{V}{}/T$ as a function of the searching time $T$ in the case of three searching strategies (averaged over all the nodes and $100$ realizations for each of them): $q\gg 1$ (blue tringles), $q=q_p$ (red squares) and the adaptive strategy (black circles). Bottom panel: mean information $\avem{I}{}$ as a function of $T$, again for the three best searching strategies (represented with the same colors as above). Error bars represent the dispersion (standard deviation) among the values obtained for different home-nodes.}
\label{fig6a}
\end{center}
\end{figure}

Figure\ \ref{fig6a} shows results for the final efficiency $E(T)$ and the information gathered $I(T)$ for the three best strategies: 
the adaptive one, $q\to \infty$ and $q=q_p$ (although this is not really a \ti{strategy} since we need to know the precise value of $q_p$). Both quantities confirm that, unless $T$ is more than twice the network size $N=10000$, the best performance is obtained for $q=q_p$. Nevertheless, our adaptive strategy gives results that are very close to those obtained for $q=q_p$ and always better than those obtained for $q\gg 1$ (at least for $T\leq2N$) both in terms of efficiency and in terms of the total amount of information recovered.

All these results are coherent with the description of the walkers' behavior commented on in the previous section. In particular, it is reasonable that when $q \gg 1$ the efficiency initially increases with $T$ since in this case the shorter the searching duration, the larger the probability that an agent gets lost. On the contrary, for $q_p$ and the adaptive strategy, which precisely aims at capturing the behavior of the agents at $q_p$, the information is mainly collected by means of quite short round trips. Consequently, increasing the searching time reduces the efficiency because it increases the chance to visit many times the same nodes. In any case, when $T \gg N$ and $I \sim 1$, the problem of visiting already visited nodes becomes relevant also for the strategy  $q\gg 1$. Finally, it is worth stressing that while for $q \gg 1$, the dispersion among the values $E^{(i)}$ and $I^{(i)}$ for different home-nodes is very high, in the case of the other two strategies, the same does not happen. This is a clear indication of the fact that the adaptive strategy recovers one of the most interesting features of the agents' behavior at $q_p$, namely, the homogeneity of the performance starting from different home-nodes. We next discuss one potential application of the searching strategies previously discussed. This would also allow for a better distinction of what strategy is the best. 

\begin{figure*}[thbf]
\begin{center}
\includegraphics[width=.8\linewidth]{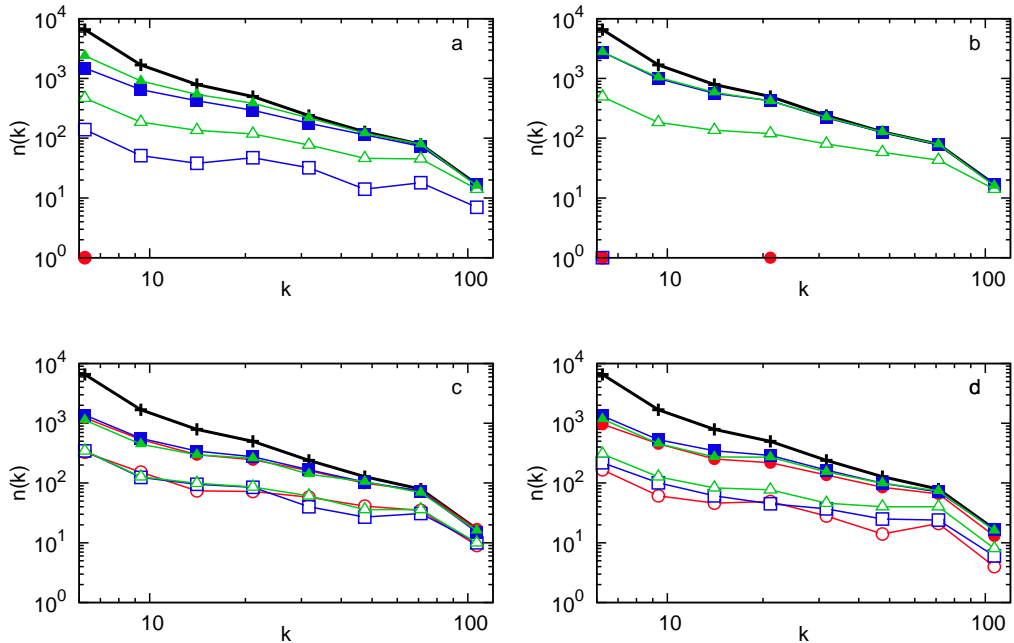}
\caption{(Color online). Number of nodes of degree $k$ as measured by a walker searching during $T=2000$ (void symbols) or $T=10000$ (filled symbols) time steps, in a single realization of the process. Red circles refers to a home-node of degree $k=5$, blue squares to a home-node with degree $k=22$ and green triangles to the node with the largest degree in the network ($k=100$). The searching strategies are, respectively: $q=1$ in panel (a), $q\gg1$ in (b), $q=q_p$ in (c) and the adaptive strategy in (d). The black line is the real degree distribution.}
\label{fig7a}
\end{center}
\end{figure*}


\begin{figure*}[tb]
\begin{center}
\includegraphics[width=.9\linewidth]{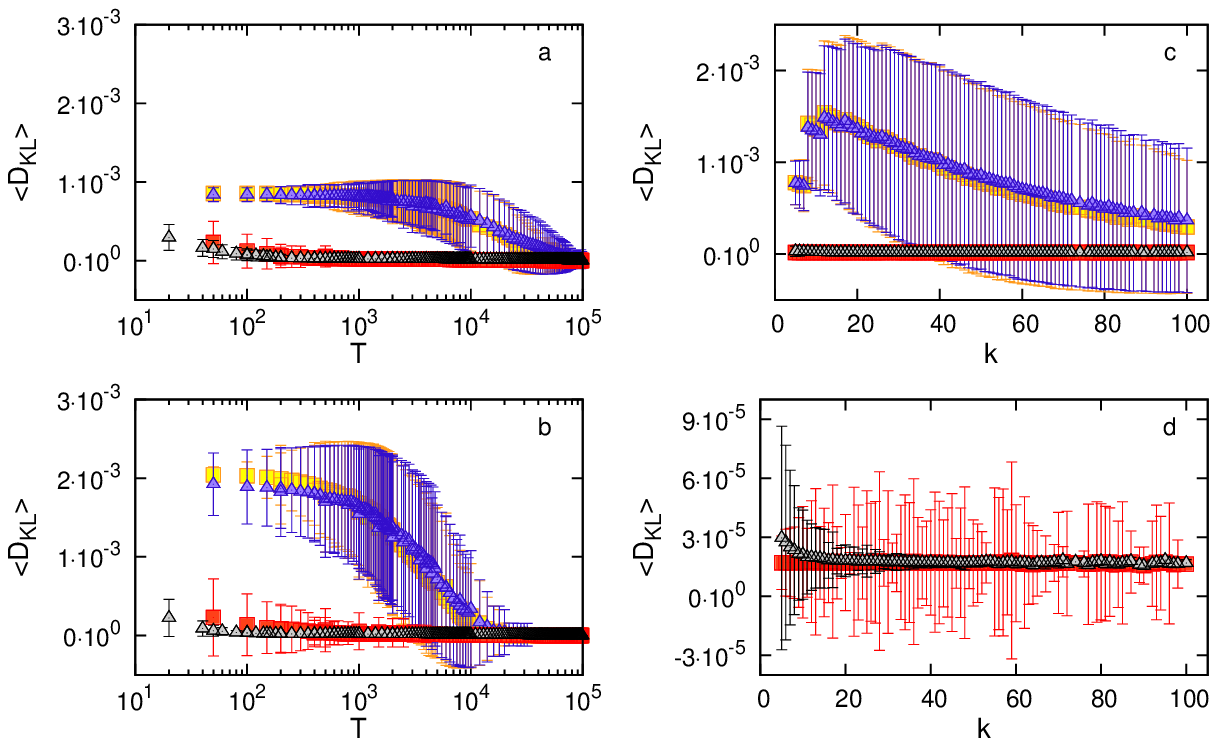}
\caption{(Color online). Kullback Divergence ($D_\mathrm{KL}$) of the measured (normalized) degree distribution with respect to the real one. Different colors refer to different searching strategies: orange squares (top curve) for $q\gg1$, blue triangles (top curve) for $q=1$ (pure random walk), red squares (bottom curve) for $q=q_p$ and black triangles (bottom curve) for the adaptive strategy. In panels (a)-(b) $D_\mathrm{KL}$ is plotted as a function of the searching duration $T$, averaged over $10^4$ realizations. The considered home-node are, in panel (a), the node with the largest degree ($k=100$) and, in panel (b), a node with degree $k=5$.
In panels (c)-(d), $D_\mathrm{KL}$ is plotted as a function of the degree of the home-node, averaged over $[1/p(k_H)]$ realizations for each starting node of degree $k_H$. The searching time is $T=2000$. In panel (d) the adaptive strategy is plotted on a smaller scale together with the strategy $q=q_p$.}
\label{fig8a}
\end{center}
\end{figure*}

\section{Recovering the degree distribution.}
\label{section6}

An important global descriptor of every network is its degree distribution $P(k)$. However, this information is not always at hand. For instance, suppose you belong to a network of which you only know your local neighborhood (like an online social network or a city map). The problem is then to know what is your position in the network as far as the degree is concerned or to make an exploration that allow you to gather information about the entire map. In other words, we want to study if the sample of nodes visited by an agent is more or less representative of the global system, at least with regard to its $P(k)$.

In Fig.\,\ref{fig7a} we plot the number of nodes of degree $k$, $N(k)$, found in a typical realization of the different strategies, for two different values of $T$ and for different choices of home-nodes. As we expected, the usual random walker and the agent with $q \gg 1$ are very bad when the home-node has a small degree (red curves). On the contrary, the performance of the adaptive protocol and that of the walker when $q=q_p$ are almost not affected by the walkers' lifetimes (at least for the considered values) and by the degree of the home-node. Note that panels (a) and (b) represent the common situation in which a random walker starting at a lowly connected home node gets lost. Indeed, for such cases, the only information brought back is the degree of the node from which the walker started the exploration of the network. However, as it is also appreciated in the figure, when the home node has a relative high degree, setting $q \gg 1$ constitutes the best strategy for an accurate estimation of $N(k)$. Nevertheless, as the walker "does not know" what is the connectivity of its home node in relation to the rest of the network, the latter strategy seems to be, as a rule of thumb, a bad choice. Additionally, the figure also shows that in general what is difficult for an agent to recover are the most peripheral nodes of the network. Consequently, the nodes with a small degree are usually under represented while the heavy tail of the degree distribution is reconstructed with high accuracy. 

In order to quantify the accuracy of the reconstructed networks, we calculate the Kullback-Leibler Divergence or relative entropy \cite{kl51}, a non-symmetric measure of the difference between two probability distributions. This is a standard method to evaluate how different an experimentally estimated distribution is from the real one. For the probability distributions $P$ and $Q$ of a discrete random variable their K\,L divergence is defined to be
\begin{equation}
D_{\mathrm{KL}}(P\|Q) = \sum_k P(k) \log \frac{P(k)}{Q(k)},
\end{equation}
where $P(k)$ is the real distribution and $Q(k)$ the estimated one. Using this measure, we explore how the accuracy of the reconstruction depends on the searching strategy, the walkers' lifetimes and the degree of the home node. In what follows, we report results for the mean values, averaged over many realizations, and for the deviations around the means.

In Fig.\,\ref{fig8a}, panels (a)-(b), we plot $D_\mathrm{KL}$ as a function of $T$ for four different strategies (including the standard random walk) and two different starting nodes (corresponding to, respectively, maximum and minimum degree,). As expected, $D_\mathrm{KL} \to 0 $ when $T\to \infty$, in all the considered cases. 
The $q=q_p$ strategy and the adaptive protocol perform much better than the other two settings, with less dispersion and a very much weaker dependence on the degree of the home node. Hence, these last two strategies are more suited if we aim at recovering $P(k)$, especially when $T$ is not too long. Moreover, even if they both are good, the adaptive strategy is better than $q=q_p$, with a very small dispersion. Thus, although it is not possible to perform better that $q=q_p$ in terms of nodes visited, the adaptive strategy does better in terms of the accuracy of $Q(k)$, that is, when it comes to reconstruct $P(k)$.  

We have also analyzed the dependency of $D_\mathrm{KL}$ on the degree of the home nodes for fixed values of $T$. Figure\,\ref{fig8a}, panels (c)-(d), displays $D_\mathrm{KL}$ as a function of $k$ for all the strategies, in the case of a short lifetime ($T=2000$). Differences among strategies are really noteworthy, while for the larger lifetime ($T=10000$) we verified that they persist just in the case of quite small degrees of the home nodes. Finally, the adaptive strategy is in general the best option, being $q=q_p$ slightly better only in the case of home nodes with degree $k<\langle k\rangle$.

\section{Conclusions}
\label{section7}

In this paper, we have presented a model for network search and exploration in which walkers evaluate at each time step whether to go farther from a home node or get back with the information retrieved up to that moment. These probabilities depend on a single parameter $q$, that has been shown to exhibit an optimal value, $q=q_p<1$ ($q=1$ corresponds to the markovian random walk limit) for exploration times comparable to the system size. When the walkers are allowed to explore the network indefinitely or during long times, the optimal value turns out to be $q=\infty$. However, although the amount of information recovered for the latter choice could be maximal, the results are highly dependent on the degree of the home node: the smaller the degree of the node assigned to the walker, the less information the walker can get back home. As a matter of fact, for most of the nodes (recall that in a scale-free network most of the nodes are poorly connected), $q=\infty$ is not the best strategy.

Capitalizing on the behavior of the walkers as a function of $q$, we have also proposed an alternative algorithm in which the agents are allowed to tune the value of the parameter $q$ to optimize the information retrieved. Through numerical simulations, we have shown that this mechanism allows an exploration as efficient as that performed setting $q=q_p$. Nevertheless, the adaptive scheme has the advantage that the value of $q$ is changed dynamically, and therefore it overcomes the problem of fixing an a priori unknown optimal value $q_p$. We believe that this adaptive search protocol could be a valuable addition to the current literature as it performs optimally with a minimum (local) information about the network structure. 

As a demonstration of the potentialities of the algorithms explored in this work, we have made use of the different searching strategies to address the problem of network discovery. As expected, the adaptive mechanism is the one whose performance, in terms of the quality and quantity of the information retrieved, is the best. Whether or not these kinds of strategies can be further developed and applied to the exploration of real networks is out of the scope of the present paper, but we identify at least two scenarios in which they can be useful: the discovery of new connections in communication networks and the exploration of planar networks (i.e., city networks) using minimal local information. We therefore hope that our work guide future research along these lines. 

\begin{acknowledgments} 
The work has received financial support from Spanish MICINN  (grants  FIS2009-13730 and FIS2011-25167), from Generalitat de Catalunya (2009SGR00838), from the FET-Open project
DYNANETS (grant no. 233847) funded by the European Commission, and from Comunidad de Arag\'on (FMI22/10). L.P. was supported by the Generalitat de Catalunya through the FI Program
and also acknowledges hospitality of ISI Torino, where part of this work was performed.

\end{acknowledgments}


\end{document}